\documentclass[twocolumn,showpacs,preprintnumbers,amsmath,amssymb,pra,superscriptaddress]{revtex4}

\usepackage{graphicx}
\usepackage{dcolumn}
\usepackage{bm}

\begin{document}

\preprint{APS/123-QED}

\title{A spectroscopic study of the cycling transition $4s[3/2]_2-4p[5/2]_3$ at 811.8 nm in $\rm{^{39}Ar}$: \\ Hyperfine structure and isotope shift }

\author{W. Williams}
\email{wwilliams@phy.anl.gov}
\affiliation{Physics Division, Argonne National Laboratory, Argonne, Illinois 60439, USA}
\author{Z.-T. Lu}
\affiliation{Physics Division, Argonne National Laboratory, Argonne, Illinois 60439, USA}
\affiliation{Department of Physics and Enrico Fermi Institute, University of Chicago, Chicago, Illinois 60637, USA}
\author{K. Rudinger}
\affiliation{Physics Division, Argonne National Laboratory, Argonne, Illinois 60439, USA}
\affiliation{Department of Physics and Enrico Fermi Institute, University of Chicago, Chicago, Illinois 60637, USA}
\author{C.-Y. Xu}
\affiliation{Physics Division, Argonne National Laboratory, Argonne, Illinois 60439, USA}
\affiliation{Department of Physics and Enrico Fermi Institute, University of Chicago, Chicago, Illinois 60637, USA}
\author{R. Yokochi}
\affiliation{Department of Earth and Environmental Sciences, University of Illinois at Chicago, Chicago, Illinois 60607, USA}
\author{P. Mueller}
\email{pmueller@anl.gov}
\affiliation{Physics Division, Argonne National Laboratory, Argonne, Illinois 60439, USA}
\date{\today}

\begin{abstract}
Doppler-free saturated absorption spectroscopy is performed on an enriched radioactive $\rm{^{39}Ar}$ sample. The spectrum of the $3s^2 3p^5 4s [3/2]_2 - 3s^2 3p^5 4p [5/2]_3$ cycling transition at 811.8 nm is recorded, and its isotope shift between $\rm{^{39}Ar}$ and $\rm{^{40}Ar}$ is derived. The hyperfine coupling constants A and B for both the $4s [3/2]_2$ and $4p [5/2]_3$ energy levels in $\rm{^{39}Ar}$ are also determined. The results partially disagree with a recently published measurement of the same transition. Based on earlier measurements as well as the current work, the isotope shift and hyperfine structure of the corresponding transition in $\rm{^{37}Ar}$ are also calculated. These spectroscopic data are essential for the realization of laser trapping and cooling of $\rm{^{37,39}Ar}$.
\end{abstract}

\maketitle

\section{\label{sec:level1}Introduction}
\begin{figure}
{\includegraphics[width=0.9\linewidth]{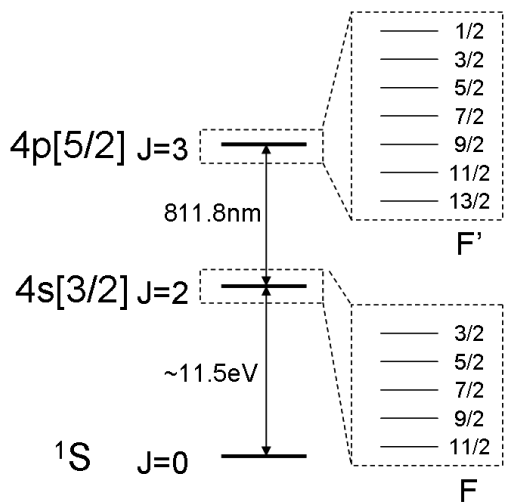}}
\caption{\label{fig:figure1} Simplified energy level diagram of argon including the hyperfine levels for the spin 7/2 isotope $^{39}$Ar. $\rm{4s}\mbox{-}\rm{4p}$ ($\rm{F=11/2-F'=13/2}$) is the ``cycling transition'' to be used for laser trapping and cooling of metastable argon atoms.}
\end{figure}
Trace analysis of the long lived noble gas radionuclides $\rm{^{85}Kr}$ ($t_{1/2}=10.75~\rm{years}$, isotopic abundance $\rm{^{85}Kr/Kr} =10^{-12}$), $\rm{^{81}Kr}$ ($t_{1/2}=230,000~\rm{years}$, $\rm{^{81}Kr/Kr}\sim 10^{-13}$), and $\rm{^{39}Ar}$ ($t_{1/2}=269~\rm{years}$, $\rm{^{39}Ar/Ar}\sim 8 \times 10^{-16}$) can be used to determine the mean residence time, so called ``age'', of old ice and groundwater samples \cite{Collon_2004}. While $^{85}$Kr is an anthropogenic isotope resulting from nuclear fission, $^{81}$Kr and $^{39}$Ar are predominantly cosmogenic isotopes, i.e., they are produced in the atmosphere through cosmic radiation induced nuclear reactions on stable krypton and argon, respectively. The three tracer isotopes have very different half-lives, each can be used to cover a different age range. This is of great interest in many areas of earth science research including the study of climate history, hydrology, glaciology, and oceanography. There are currently three methods of trace analysis that are sensitive to these noble gas isotopes at the isotopic abundance level of parts-per-trillion: Low-level counting (LLC), accelerator mass spectrometry (AMS), and atom trap trace analysis (ATTA).  The capabilities of these three methods, their relative advantages and disadvantages, have been reviewed in \cite{Collon_2004}. ATTA is the newest method among the three. In ATTA, individual atoms of the desired isotope are captured by a magneto-optical trap and detected by monitoring their fluorescence \cite{Chen_1999}. The superb selectivity and sensitivity of ATTA has been demonstrated with the analyses of $\rm{^{81}Kr}$ and $\rm{^{85}Kr}$ in reference and environmental samples \cite{Du_2004,Sturchio_2004}. In principle, ATTA can also be used to analyze $\rm{^{39}Ar}$. However, the frequencies of the cycling $\rm{4s}\mbox{-}\rm{4p}$ transition in $\rm{^{39}Ar}$ (see Figure 1 for an energy level diagram) must first be determined with an uncertainty of $\sim1~\rm{MHz}$ before it is possible to realize a magneto-optical trap or ATTA of this isotope. Since the natural abundance of $\rm{^{39}Ar}$ is extremely small, such spectroscopic studies are best carried out using samples highly enriched in this radioactive isotope.

While $\rm{^{39}Ar}$ has been studied by Klein \em et al. \rm \cite{Klein_1996} using high-resolution laser spectroscopy, precision data on the cycling transition have been incomplete until recently \cite{Welte_2009}. The magnetic dipole and the electric quadrupole hyperfine coupling constants, $A$ and $B$, had been measured for the $4s[3/2]_2$ level of $\rm{^{39}Ar}$ together with isotope shifts of the $4s[3/2]_2-4p[3/2]_2$ ($764~\rm{nm}$) transition at the CERN/ISOLDE on-line isotope production facility using high resolution collinear laser spectroscopy \cite{Klein_1996}. Isotope shifts of the cycling transition were measured for the three stable isotopes $\rm{^{36}Ar}$, $\rm{^{38}Ar}$ and $\rm{^{40}Ar}$ reaching $\sim1~\rm{MHz}$ resolution  \cite{Damico_1999}. Recently, isotope shifts and hypefine structure of the cycling transition have also been measured for $^{39}$Ar using modulation-transfer saturated absorption spectroscopy in an enriched sample \cite{Welte_2009}. In the work presented here, a slightly different technique of frequency modulated (FM) saturated absorption spectroscopy \cite{Hall_1981} is performed on a radioactive $\rm{^{39}Ar}$ sample. High resolution $\rm{^{39}Ar}$ spectra of the $3s^2 3p^5 4s[3/2]_2-3s^2 3p^5 4p[5/2]_3$ cycling transition at 811.8 nm (in vacuum) are recorded. Its isotope shift and hyperfine coupling constants for both the lower and upper levels are determined. Our results significantly disagree with the findings of Welte \em et al. \rm \cite{Welte_2009} with respect to the magnetic dipole hyperfine constant $A$ of both levels. Results obtained with our apparatus on hyperfine spectroscopy on $^{83}$Kr, however, agrees very well with previous work done on that isotope \cite{Cannon_1990,Cannon_1993}. Based on earlier measurements as well as the current work, we calculate the isotope shift and hyperfine structure of the corresponding transition in $\rm{^{37}Ar}$ ($t_{1/2}=35~\rm{days}$). This isotope could be valuable as a short-lived tracer isotope for quantitative $^{39}$Ar analysis. Thes spectroscopic data are essential for the realization of laser trapping and cooling of $\rm{^{37,39}Ar}$.

\section{Experimental Setup}
\begin{figure}
{\includegraphics[width=0.8\linewidth]{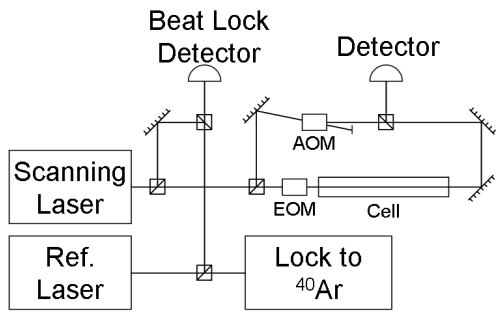}}
\caption{\label{fig:figure2} A schematic of the saturated absorption spectroscopy setup. The 1~m long spectroscopy quartz cell is filled with a mixture of an enriched $\rm{^{39}Ar}$ sample ($\rm{^{39}Ar/^{40}Ar} \sim 1.5 \times 10^{-3}$, $\sim0.5$~Torr partial pressure) and natural krypton ($\sim2$~mTorr partial pressure).}
\end{figure}
A simplified experimental layout is shown in Figure \ref{fig:figure2}.  There are two tunable, grating stabilized diode lasers: the reference laser is locked to the cycling transition in the stable and abundant $\rm{^{40}Ar}$ via basic saturated absorption spectroscopy in a separate gas cell, and the scanning laser is used for FM saturated absorption spectroscopy of $\rm{^{39}Ar}$. The frequency of the scanning laser is controlled by a beat lock with the reference laser with a standard deviation of $\sim 20~\rm{kHz}$ in the beat frequency.  This setup provides frequency control of the scanning laser within $\pm 1.5~\rm{GHz}$ relative to the reference laser. In the FM saturated absorption spectroscopy scheme, the probe beam passes through an electro-optical modulator (EOM) and is phase modulated at $9.25~\rm{MHz}$.  The pump beam passes through an accusto optical modulator (AOM) driven at $50~\rm{MHz}$ with a $90~\rm{kHz}$ square wave amplitude envelope and is overlapped with the probe beam in the spectroscopy cell. The probe beam is detected with a high bandwidth (100 MHz) photoreceiver and the resulting signal is demodulated at $9.25~\rm{MHz}$ using a radiofrequency mixer and at $90~\rm{kHz}$ using a lock-in amplifier. This scheme provides Doppler-free spectroscopy signals with a signal-to-noise ratio close to the photon shot noise limit.

The enriched $\rm{^{39}Ar}$ sample is produced by irradiating a pellet of $2~\rm{g}$ of potassium fluoride (KF) with high energy neutrons (up to $8~\rm{MeV}$, $5\times 10^{13}~\rm{s^{-1} cm^{-2}}$ total flux) in the cadmium-lined in-core radiation tube of the Oregon State University Radiation Center research reactor. The $\rm{^{39}K(n, p) ^{39}Ar}$ nuclear reaction has a cross-section of $\sim 0.3~\rm{barn}$ above neutron energies of $\sim 4~\rm{MeV}$ \cite{ENDF_2006}. After one hour of irradiation at a reactor power of 1 MW thermal, the amount of $\rm{^{39}Ar}$ contained in the sample is calculated to be $\sim 300~\rm{nCi}$ or $1.3\times 10^{14}$ atoms. After irradiation, the $\rm{^{39}Ar}$ is extracted out of the the bulk KF sample by melting the pellet at $900~^{\circ}\rm{C}$ under vacuum in a dedicated extraction apparatus. Chemically active gases in the sample overhead are removed by a getter pump. Subsequently, the residual gases, dominated by argon from air trapped in the KF sample, are frozen into a liquid nitrogen cooled charcoal absorber. The getter pump is then reactivated and the frozen gas sample is released into the vacuum system for further purification. The purity of the gas sample is monitored by a residual gas analyser (RGA) attached to the extraction vacuum system. The entire extraction process is about $80\%$ efficient. The processed gas sample contains $\sim 1 \times 10^{14}$ $\rm{^{39}Ar}$ atoms ($240~\rm{nCi}$) at an isotopic abundance of $\rm{^{39}Ar/^{40}Ar} \sim 1.5 \times 10^{-3}$.
\begin{figure*}[t]
{\includegraphics[width=0.8\linewidth]{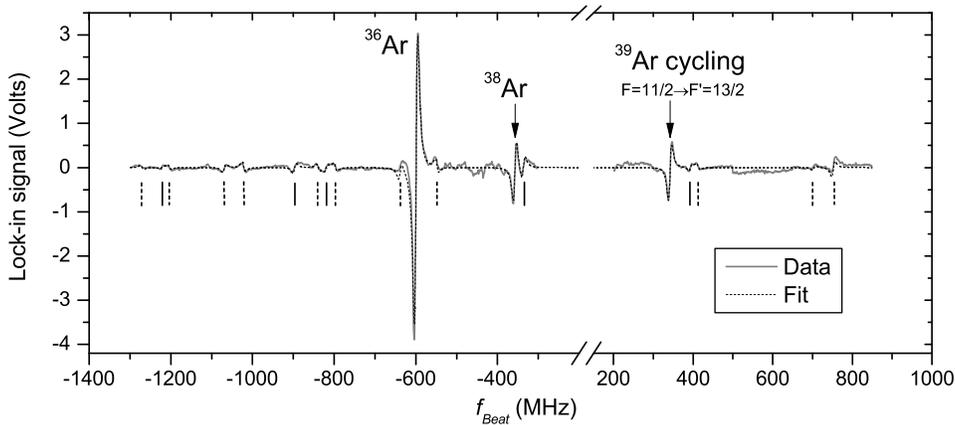}}
\caption{\label{fig:figure3} FM saturated absorption spectroscopy data of the $4s[3/2]_2-4p[5/2]_3$ transition in $\rm{^{39}Ar}$ is plotted as a solid line.  The frequency range from $f_{Beat} = -300~\rm{MHz} ~\rm{to}~ 200~\rm{MHz}$ is excluded from the scan because of the large noise introduced by the absorption signal of the abundant $\rm{^{40}Ar}$. The absorption peaks of the two minor stable isotopes $^{36}$Ar and $^{38}$Ar and of the cycling transition in the radioactive $^{39}$Ar exhibit the expected dispersion type signal shape.  Additional hyperfine transitions and cross-over signals of $^{39}$Ar identified in the data are indicated by the vertical solid and dashed markers, respectively. The dotted line shows a global fit to the data.}
\end{figure*}

Laser trapping of argon atoms is realized by exciting the $4s[3/2]_2-4p[5/2]_3$ cycling transition at 811.8 nm. The lower $\rm{4s}$ level is $11.5~\rm{eV}$ above the ground state, and is metastable with a vacuum lifetime of approximately $40~\rm{s}$ \cite{Katori_1993}. In this work, an RF-driven gas discharge is used to populate the atoms into the metastable level. The spectroscopy cell is made out of quartz and is $1~\rm{meter}$ long and $2.5~\rm{cm}$ in diameter. Since the argon sample size is too small to sustain a discharge in our vacuum system, krypton carrier gas is added to the cell to reach an operation pressure of $\sim 0.5~\rm{mTorr}$. This pressure is selected to maximize the metastable population of argon \cite{Rudinger_2009}. It should be noted that the discharge generates a plasma sheath near the cell walls that efficiently implants ions into the glass walls, causing sample loss \cite{Butler_1963,Cavaleru_1972}. If starting from a clean spectroscopy cell, over $90\%$ of the gas in an initial fill at $2~\rm{mTorr}$ is implanted into the cell walls within a few seconds of running the discharge. We found that the implantation loss can be significantly reduced by treating the cell with a krypton discharge for $\sim 12$ hours at a cell temperature of $\sim 200~^{\circ}\rm{C}$ prior to filling the cell with the enriched sample. Following such treatment, an initial fill at $2~\rm{mTorr}$ of krypton results in a stable operating pressure of $\sim0.5$~mTorr in the cell when the discharge is on. Under these conditions, spectroscopic signals of argon can be observed over many hours.

\section{Results and Discussion}

The FM saturated absorption spectrum (i.e., the output voltage of the lock-in amplifier) is recorded with the scanning laser advancing at $1~\rm{MHz}$ steps every $0.375~\rm{s}$, completing a $400~\rm{MHz}$ scan in $2.5~\rm{minutes}$.  The scans are performed in both directions, beat frequency ($f_{Beat}$) increasing and decreasing, to explore any systematic shifts, e.g. due to integration delays by the lock-in amplifier. Spectra centered on $\rm{^{36}Ar}$ ($0.3\%$ natural abundance) and $\rm{^{38}Ar}$ ($0.06\%$) are also periodically scanned as a systematic check. A sample spectrum recorded over a frequency range of 2.5~GHz is shown in Figure \ref{fig:figure3}. Absorption peaks in FM modulated spectroscopy are detected by measuring the relative phase of the probe light modulation and appear as typical dispersion-shaped features in the spectrum. Signal in the frequency range of $f_{Beat} = -300~\rm{MHz}$ to $200~\rm{MHz}$ is not shown in Figure \ref{fig:figure3} and exluded from the analysis due to a large increase in noise caused by the dominant signal of the abundant $\rm{^{40}Ar}$ ($\sim 99.6\%$ natural abundance). Outside of this frequency range we did not observe any systematic frequency shifts of the fitted peak positions that might have resulted from the $^{40}$Ar peak tails.
\begin{table}[b]
\caption{\label{tab:table1}Hyperfine constants of $\rm{^{39}Ar}$ as measured in this work and experimental values from \cite{Welte_2009}$^a$ and \cite{Klein_1996}$^b$. The respective hyperfine constants for $^{37}$Ar are either taken directly from measured values in \cite{Klein_1996}$^b$ or are calculated based on this work$^c$ (see text).}
\begin{ruledtabular}
\begin{tabular}{cccc}
& $\rm{^{39}Ar}(MHz)$  & $\rm{^{39}Ar}(MHz)$ & $\rm{^{37}Ar}(MHz)$ \\
& This work & Other works & \\
\hline
$A(4s[3/2]_2)$  & -285.36(13) & $-287.15(14)^a$ & \\
& & $-286.1(14)^b$ & $482.1(3)^b$ \\
$B(4s[3/2]_2)$ & 117.8(12) & $119.3(15)^a$ & \\
& & $118(20)^b$ & $-77.0(16)^b$\\
$A(4p[5/2]_3)$ & -134.36(8) & $-135.16(12)^a$ & $227.0(2)^c$\\
$B(4p[5/2]_3)$ & 113.1(14) & $113.6(19)^a$ & $-73.9(19)^c$\\
\end{tabular}
\end{ruledtabular}
\end{table}

The two tallest peaks at negative beat frequencies are the absorption signals from the stable isotopes $^{36}$Ar and $^{38}$Ar as indicated. These signals are also observed in scans of natural argon samples, their amplitude ratio matches well with the natural $^{36}$Ar/$^{38}$Ar abundance ratio of 5:1, and the center frequencies agree with the previously measured isotope shifts (see Table \ref{tab:table4}). The tallest peak on the positive side of $\rm{^{40}Ar}$, with beat frequency $f_{Beat}\sim 341~\rm{MHz}$ ($941.76(10)~\rm{MHz}$ above $\rm{^{36}Ar}$), is only observed in the enriched sample and is identified as the cycling transition $\rm{F}=11/2\rightarrow 13/2$ in $\rm{^{39}Ar}$. This is the transition that needs to be excited for laser trapping and cooling of $\rm{^{39}Ar}$. Its observed signal amplitude correlates well with the isotopic ratio of $\rm{^{39}Ar/^{40}Ar} \sim 1.5 \times 10^{-3}$ that we expect for the radioactive sample. This ratio is roughly a factor of two higher than the natural $^{38}$Ar abundance. However, the amplitude of the absorption signals of $^{39}$Ar are reduced by hyperfine splitting and influenced by optical pumping.
In addition to the strong cycling transition, a total of sixteen hyperfine transitions as well as both positive and negative cross-over peaks in $^{39}$Ar are observed above noise over the whole spectrum. Cross-over signals are a typical feature of saturated absorpion spectroscopy and appear when two transitions connected by a common energy level overlap within the Doppler width \cite{Demtroeder_2008}. They occur exactly at the mean frequency of the two respective transitions and can lead to enhanced or reduced transmission of the probe beam. In Figure \ref{fig:figure3} the frequency positions of the regular hyperfine transitions are indicated by the vertical solid lines, while the crossover transitions are marked by the dashed lines.
\begin{table}[b]
\caption{\label{tab:table2}Hyperfine constants and isotope shift for $\rm{^{83}Kr}$ from this work and from previous experimental work in \cite{Cannon_1993}$^a$ and \cite{Cannon_1990}$^b$.}
\begin{ruledtabular}
\begin{tabular}{ccc}
& $\rm{^{83}Kr}$(MHz) & $\rm{^{83}Kr}$(MHz) \\
& This work & Other works \\
\hline
$A(4s[3/2]_2)$  & -244.04(3) & $-243.87(5)^a$, $-243.93(4)^b$ \\
$B(4s[3/2]_2)$ & -453.36(59) & $-453.1(7)^a$, $-452.93(60)^b$\\
$A(4p[5/2]_3)$ & -103.91(2) & $-103.73(7)^a$, $-104.02(6)^b$ \\
$B(4p[5/2]_3)$ & -437.03(67) & $-438.8(12)^a$,$-436.9(17)^b$\\
$\Delta \nu^{78,83}$ & 201.33(13) & $201.9(25)^b$\\
\end{tabular}
\end{ruledtabular}
\end{table}

For analysis, global fits to the complete data are performed to determine the isotope shift of the transition, as well as the hyperfine constants for both the 4s and 4p levels.  In the analysis, we first use the sum of a Lorentzian peak shape and the derivative of a Lorentzian to fit the slightly asymmetric $\rm{^{36}Ar}$ peak and to generate a peak template for the global fit. Relative amplitude and width of these line shapes are then fixed and used in the global fit for all peaks. As mentioned above, scans of increasing frequency are separated from scans of decreasing frequency in an attempt to isolate systematic errors. The hyperfine constants determined from the scans in both directions are consistent within error bars.  On the other hand, the isotope shifts determined from the scans in opposite directions differ by $\sim 540~\rm{kHz}$, about three times the statistical error.  This difference is probably due to delays in the detector-amplifier system. For a conservative error estimate, we take the standard deviation of the measured isotope shift values as the respective systematic error due to this effect. The values from the global fits as well as the previously reported values are shown in Table \ref{tab:table1}.

Obviously, our results for the magnetic dipole hyperfine constant $A$ disagree significantly with the values from Welte \em et al. \rm \cite{Welte_2009} for both levels involved, while the values for the electric quadrupole constant $B$ generally agree within the errors. Both measurements claim on the order of $\sim 100$~kHz uncertainties in the determination of $A$, but the results differ by one to two MHz. The less precise value from Klein \em et al. \rm \cite{Klein_1996} for the $A$ of the lower level agrees with both newer measurements and can not resolve the discrepancy. Interestingly, the ratio $A(4s[3/2]_2)/A(4p[5/2]_3)$ from this work agrees very well with that from \cite{Welte_2009} to within $<0.1\%$ ($2.124(2)$ \em vs. \rm $2.125(2)$).

To check our results for consistency, FM saturation spectroscopy is also performed on the $5s[3/2]_2-5p[5/2]_3$ transition ($811.5$~nm) in a natural krypton sample, which includes the stable $\rm{^{83}Kr}$ (nuclear spin = 9/2, isotopic abundance $11.5\%$). This measurement was carried out with the identical experimental apparatus and under very similar experimental conditions with respect to discharge and laser power settings. It serves as a systematic test of the measurement and analysis procedure and specifically confirms that the tallest peak indicates the cycling transition. The results, which provide a slight improvement over previous knowlede of this transition \cite{Cannon_1990,Cannon_1993}, are shown in Table \ref{tab:table2}. Generally, the results agree well with the literature values at the $100 - 200~\rm{kHz}$ level for $A$ and $\sim 1~\rm{MHz}$ for $B$, which is consistent with the stated uncertainties of our $^{39}$Ar results.

Combining the results of this work on $\rm{^{39}Ar}$ and previous results on different transitions in $\rm{^{37}Ar}$ ($t_{1/2}\sim 30~\rm{days}$), the frequencies of the $\rm{4s}\mbox{-}\rm{4p}$ cycling transition in $\rm{^{37}Ar}$ can also be derived.  The hyperfine constants for the $4s[3/2]_2$ level of $\rm{^{37}Ar}$ were measured by Klein \em et al. \rm \cite{Klein_1996} (see Table \ref{tab:table1}).  Hyperfine anomalies due to differences in the distribution of the magnetic dipole density between $\rm{^{37}Ar}$ and $\rm{^{39}Ar}$ are predicted to influence the determination of the hyperfine constants only at a relative level of $\sim 10^{-4}$ \cite{Klein_1996}.  Ignoring this small correction here, we can calculate A for the 4p level of $\rm{^{37}Ar}$ by equating the ratio of hyperfine constants:
\begin{equation}
\frac{A_{39}(4s[3/2]_2)}{A_{39}(4p[5/2]_3)}=\frac{A_{37}(4s[3/2]_2)}{A_{37}(4p[5/2]_3)}\
\end{equation}
A similar ratio can be used to calculate the B hyperfine coefficient. The isotope shift of the $4s[3/2]_2-4p[5/2]_3$ transition for $\rm{^{37}Ar}$ can be calculated from the reported nuclear charge radius of $\rm{^{37}Ar}$ \cite{Klein_1996} and the specific mass shift and field shift calculated from the isotope shifts and charge radii of $\rm{^{36}Ar}$, $\rm{^{38}Ar}$ and $\rm{^{40}Ar}$ for this transition \cite{Damico_1999}.

The isotope shift $\Delta \nu^{A,A^\prime}$ between two isotopes with mass numbers $A$ and $A^\prime$ can in first order be calculated from the equation
\begin{equation}
\Delta \nu^{A,A^\prime}=(K_{NMS}+K_{SMS}) \frac{A^\prime-A}{AA^\prime}+K_{KS} ~ \delta \! \left\langle r^2 \right\rangle^{A,A^\prime}
\end{equation}
where $K_{NMS}$ and $K_{SMS}$ are the normal and specific mass shift constants, respectively, $K_{FS}$ is the field shift constant, and $\delta \! \left\langle r^2 \right\rangle ^{A,A^\prime}$ is the change in the mean square charge radius between the respective isotopes \cite{Otten_1989}. While $K_{NMS}$ is trivially calculated from the transition frequency, $K_{SMS}$ and $K_{FS}$ are determined from experimental data.

\begin{table}[th]
\caption{\label{tab:table4}Isotope shift of the $4s-4p$ transition and changes in mean square charge radii relative to $^{38}$Ar: $^a$this work experimental,  $^b$this work calculation (see text), $^c$from \cite{Damico_1999}, $^d$adapted from \cite{Welte_2009}, $^e$adapted from Klein \em et~al. \rm \cite{Klein_1996}.}
\begin{ruledtabular}
\begin{tabular}{ccc}
 $A^\prime$ & $\rm{\Delta \nu^{38,A^\prime}}$ (MHz) & $\delta \left\langle r^2 \right\rangle^{38,A^\prime} (fm^{2})$ \\
\hline
$\rm{^{36}Ar}$  & -242.23(25)$^a$,  -242.0(12)$^c$ & -0.082(26)$^e$ \\
$\rm{^{37}Ar}$ & -113.0(27)$^b$ & -0.081(19)$^e$ \\
$\rm{^{38}Ar}$ & 0 & 0 \\
$\rm{^{39}Ar}$ & 113.92(41)$^a$, $112.9(10)^d$ & 0.023(17)$^b$, 0.044(68)$^e$ \\
$\rm{^{40}Ar}$ & 207.9(9)$^a$ & 0.169(33)$^e$\\
\end{tabular}
\end{ruledtabular}
\end{table}

To minimize error propagation due to the previously reported charge radii, the isotope shift for $\rm{^{37}Ar}$ is calculated with respect to $\rm{^{38}Ar}$. Using the values from Klein \cite{Klein_1996}, also listed in Table \ref{tab:table4}, the specific mass shift and field shift for the $4s-4p$ transition are calculated using data from $\rm{^{36}Ar}$ and $\rm{^{40}Ar}$ to be $K_{SMS}= -30500(4200)~\rm{MHz/amu}$ and $K_{FS}=-112(49)~\rm{MHz/fm^2}$.  It is then possible to determine the isotope shift of $\rm{^{37}Ar}$ and also slightly improve on the previous value of $\delta \! \left\langle r^2 \right\rangle ^{38,39}$ as listed in Table \ref{tab:table4}. The $^{39}$Ar isotope shift reported in \cite{Welte_2009} agrees well with this work. 

\section{Conclusions}
This work presents measurements of the isotope shift and hyperfine constants for the $4s[3/2]_2 \mbox{-} 4p[5/2]_3$ cycling transition in $\rm{^{39}Ar}$ and $\rm{^{37}Ar}$ with accuracies of well below the natural linewidth of 5.9~MHz. This information will allow to accurately choose laser frequencies for the cycling transitions (e.g., the $F = 11/2~\rm{to}~13/2$ transition in $^{39}$Ar) as well as for the required sidebands for hyperfine repumping to laser cool and trap these radioactive isotopes. Applying the ATTA technique to detect the ultra rare tracer isotope $^{39}$Ar has the potential of opening a number of exciting applications in earth sciences. The challenge remains to build an ATTA system for argon with sufficient efficiency to enable analysis of isotopes at the part-per-quadrillion abundance levels with practical sample sizes and counting rates.
\begin{acknowledgments}
We would like to thank the Oregon State University Radiation Center for support in generating the enriched sample. We would like to thank Kevin Bailey, John Greene and Thomas O'Connor for technical support. This work was supported by the U.S. Department of Energy, Office of Nuclear Physics under contract No. DE-AC02-06CH11357.
\end{acknowledgments}

\bibliographystyle{apsrev}

\begin{thebibliography}{15}
\expandafter\ifx\csname natexlab\endcsname\relax\def\natexlab#1{#1}\fi
\expandafter\ifx\csname bibnamefont\endcsname\relax
  \def\bibnamefont#1{#1}\fi
\expandafter\ifx\csname bibfnamefont\endcsname\relax
  \def\bibfnamefont#1{#1}\fi
\expandafter\ifx\csname citenamefont\endcsname\relax
  \def\citenamefont#1{#1}\fi
\expandafter\ifx\csname url\endcsname\relax
  \def\url#1{\texttt{#1}}\fi
\expandafter\ifx\csname urlprefix\endcsname\relax\def\urlprefix{URL }\fi
\providecommand{\bibinfo}[2]{#2}
\providecommand{\eprint}[2][]{\url{#2}}

\bibitem[{\citenamefont{Collon et~al.}(2004)\citenamefont{Collon, Kutschera,
  and Lu}}]{Collon_2004}
\bibinfo{author}{\bibfnamefont{P.}~\bibnamefont{Collon}},
  \bibinfo{author}{\bibfnamefont{W.}~\bibnamefont{Kutschera}},
  \bibnamefont{and} \bibinfo{author}{\bibfnamefont{Z.-T.} \bibnamefont{Lu}},
  \bibinfo{journal}{Annu. Rev. Nucl. Part. Sci.} \textbf{\bibinfo{volume}{54}},
  \bibinfo{pages}{39} (\bibinfo{year}{2004}).

\bibitem[{\citenamefont{Chen et~al.}(1999)\citenamefont{Chen, Li, Bailey,
  O'Connor, Young, and Lu}}]{Chen_1999}
\bibinfo{author}{\bibfnamefont{C.~Y.} \bibnamefont{Chen}},
  \bibinfo{author}{\bibfnamefont{Y.~M.} \bibnamefont{Li}},
  \bibinfo{author}{\bibfnamefont{K.}~\bibnamefont{Bailey}},
  \bibinfo{author}{\bibfnamefont{T.~P.} \bibnamefont{O'Connor}},
  \bibinfo{author}{\bibfnamefont{L.}~\bibnamefont{Young}}, \bibnamefont{and}
  \bibinfo{author}{\bibfnamefont{Z.-T.} \bibnamefont{Lu}},
  \bibinfo{journal}{Science} \textbf{\bibinfo{volume}{286}},
  \bibinfo{pages}{1139} (\bibinfo{year}{1999}).

\bibitem[{\citenamefont{Du et~al.}(2004)\citenamefont{Du, Bailey, Lu, Mueller,
  O'Connor, and Young}}]{Du_2004}
\bibinfo{author}{\bibfnamefont{X.}~\bibnamefont{Du}},
  \bibinfo{author}{\bibfnamefont{K.}~\bibnamefont{Bailey}},
  \bibinfo{author}{\bibfnamefont{Z.-T.} \bibnamefont{Lu}},
  \bibinfo{author}{\bibfnamefont{P.}~\bibnamefont{Mueller}},
  \bibinfo{author}{\bibfnamefont{T.~P.} \bibnamefont{O'Connor}},
  \bibnamefont{and} \bibinfo{author}{\bibfnamefont{L.}~\bibnamefont{Young}},
  \bibinfo{journal}{Rev. Sci. Instr.} \textbf{\bibinfo{volume}{75}},
  \bibinfo{pages}{3224} (\bibinfo{year}{2004}).

\bibitem[{\citenamefont{Sturchio et~al.}(2004)\citenamefont{Sturchio, Du,
  Purtschert, Lehmann, Sultan, Patterson, Lu, Mueller, Bailey, O'Connor
  et~al.}}]{Sturchio_2004}
\bibinfo{author}{\bibfnamefont{N.~C.} \bibnamefont{Sturchio}},
  \bibinfo{author}{\bibfnamefont{X.}~\bibnamefont{Du}},
  \bibinfo{author}{\bibfnamefont{R.}~\bibnamefont{Purtschert}},
  \bibinfo{author}{\bibfnamefont{B.~E.} \bibnamefont{Lehmann}},
  \bibinfo{author}{\bibfnamefont{M.}~\bibnamefont{Sultan}},
  \bibinfo{author}{\bibfnamefont{L.~J.} \bibnamefont{Patterson}},
  \bibinfo{author}{\bibfnamefont{Z.-T.} \bibnamefont{Lu}},
  \bibinfo{author}{\bibfnamefont{P.}~\bibnamefont{Mueller}},
  \bibinfo{author}{\bibfnamefont{K.}~\bibnamefont{Bailey}},
  \bibinfo{author}{\bibfnamefont{T.~P.} \bibnamefont{O'Connor}},
  \bibnamefont{et~al.}, \bibinfo{journal}{Geophys. Res. Lett.}
  \textbf{\bibinfo{volume}{31}}, \bibinfo{pages}{L05503}
  (\bibinfo{year}{2004}).

\bibitem[{\citenamefont{Klein et~al.}(1996)\citenamefont{Klein, Brown, Georg,
  Keim, Lievens, Neugart, Neuroth, Silverans, and Vermeeren}}]{Klein_1996}
\bibinfo{author}{\bibfnamefont{A.}~\bibnamefont{Klein}},
  \bibinfo{author}{\bibfnamefont{B.~A.} \bibnamefont{Brown}},
  \bibinfo{author}{\bibfnamefont{U.}~\bibnamefont{Georg}},
  \bibinfo{author}{\bibfnamefont{M.}~\bibnamefont{Keim}},
  \bibinfo{author}{\bibfnamefont{P.}~\bibnamefont{Lievens}},
  \bibinfo{author}{\bibfnamefont{R.}~\bibnamefont{Neugart}},
  \bibinfo{author}{\bibfnamefont{M.}~\bibnamefont{Neuroth}},
  \bibinfo{author}{\bibfnamefont{R.~E.} \bibnamefont{Silverans}},
  \bibnamefont{and}
  \bibinfo{author}{\bibfnamefont{L.}~\bibnamefont{Vermeeren}},
  \bibinfo{journal}{Nucl. Phy. A} \textbf{\bibinfo{volume}{607}},
  \bibinfo{pages}{1} (\bibinfo{year}{1996}).

\bibitem[{\citenamefont{Welte et~al.}(2009)\citenamefont{Welte, Steinke,
  Henrich, Ritterbusch, Oberthaler, Aeschbach-Hertig, Schwarz, and
  Trieloff}}]{Welte_2009}
\bibinfo{author}{\bibfnamefont{J.}~\bibnamefont{Welte}},
  \bibinfo{author}{\bibfnamefont{I.}~\bibnamefont{Steinke}},
  \bibinfo{author}{\bibfnamefont{M.}~\bibnamefont{Henrich}},
  \bibinfo{author}{\bibfnamefont{F.}~\bibnamefont{Ritterbusch}},
  \bibinfo{author}{\bibfnamefont{M.~K.} \bibnamefont{Oberthaler}},
  \bibinfo{author}{\bibfnamefont{W.}~\bibnamefont{Aeschbach-Hertig}},
  \bibinfo{author}{\bibfnamefont{W.~H.} \bibnamefont{Schwarz}},
  \bibnamefont{and} \bibinfo{author}{\bibfnamefont{M.}~\bibnamefont{Trieloff}},
  \bibinfo{journal}{Rev. Sci. Instr.} \textbf{\bibinfo{volume}{80}},
  \bibinfo{eid}{113109} (\bibinfo{year}{2009}).
  
\bibitem[{\citenamefont{D'Amico et~al.}(1999)\citenamefont{D'Amico, Pesce, and
  Sasso}}]{Damico_1999}
\bibinfo{author}{\bibfnamefont{G.}~\bibnamefont{D'Amico}},
  \bibinfo{author}{\bibfnamefont{G.}~\bibnamefont{Pesce}}, \bibnamefont{and}
  \bibinfo{author}{\bibfnamefont{A.}~\bibnamefont{Sasso}}, \bibinfo{journal}{J.
  Opt. Soc. Am. B} \textbf{\bibinfo{volume}{16}}, \bibinfo{pages}{1033}
  (\bibinfo{year}{1999}).

\bibitem[{\citenamefont{Hall et~al.}(1981)\citenamefont{Hall, Hollberg, Baer,
  and Robinson}}]{Hall_1981}
\bibinfo{author}{\bibfnamefont{J.~L.} \bibnamefont{Hall}},
  \bibinfo{author}{\bibfnamefont{L.}~\bibnamefont{Hollberg}},
  \bibinfo{author}{\bibfnamefont{T.}~\bibnamefont{Baer}}, \bibnamefont{and}
  \bibinfo{author}{\bibfnamefont{H.~G.} \bibnamefont{Robinson}},
  \bibinfo{journal}{App. Phy. Lett.} \textbf{\bibinfo{volume}{39}},
  \bibinfo{pages}{680} (\bibinfo{year}{1981}).

\bibitem[{\citenamefont{Cannon}(1993)}]{Cannon_1993}
\bibinfo{author}{\bibfnamefont{B.~D.} \bibnamefont{Cannon}},
  \bibinfo{journal}{Phys. Rev. A} \textbf{\bibinfo{volume}{47}},
  \bibinfo{pages}{1148} (\bibinfo{year}{1993}).

\bibitem[{\citenamefont{Cannon and Janik}(1990)}]{Cannon_1990}
\bibinfo{author}{\bibfnamefont{B.~D.} \bibnamefont{Cannon}} \bibnamefont{and}
  \bibinfo{author}{\bibfnamefont{G.~R.} \bibnamefont{Janik}},
  \bibinfo{journal}{Phys. Rev. A} \textbf{\bibinfo{volume}{42}},
  \bibinfo{pages}{397} (\bibinfo{year}{1990}).

\bibitem[{\citenamefont{Chadwick et~al.}(2006)\citenamefont{Chadwick, Oblo{\v
  z}insk{\' y}, Herman et~al.}}]{ENDF_2006}
\bibinfo{author}{\bibfnamefont{M.}~\bibnamefont{Chadwick}},
  \bibinfo{author}{\bibfnamefont{P.}~\bibnamefont{Oblo{\v z}insk{\' y}}},
  \bibinfo{author}{\bibfnamefont{M.}~\bibnamefont{Herman}},
  \bibnamefont{et~al.}, \bibinfo{journal}{Nuclear Data Sheets}
  \textbf{\bibinfo{volume}{107}}, \bibinfo{pages}{2931} (\bibinfo{year}{2006}).

\bibitem[{\citenamefont{Katori and Shimizu}(1993)}]{Katori_1993}
\bibinfo{author}{\bibfnamefont{H.}~\bibnamefont{Katori}} \bibnamefont{and}
  \bibinfo{author}{\bibfnamefont{F.}~\bibnamefont{Shimizu}},
  \bibinfo{journal}{Phys. Rev. Lett.} \textbf{\bibinfo{volume}{70}},
  \bibinfo{pages}{3545} (\bibinfo{year}{1993}).

\bibitem[{\citenamefont{Rudinger et~al.}(2009)\citenamefont{Rudinger, Lu, and
  Mueller}}]{Rudinger_2009}
\bibinfo{author}{\bibfnamefont{K.}~\bibnamefont{Rudinger}},
  \bibinfo{author}{\bibfnamefont{Z.-T.} \bibnamefont{Lu}}, \bibnamefont{and}
  \bibinfo{author}{\bibfnamefont{P.}~\bibnamefont{Mueller}},
  \bibinfo{journal}{Rev. Sci. Instr.} \textbf{\bibinfo{volume}{80}},
  \bibinfo{eid}{036105} (pages~\bibinfo{numpages}{2}) (\bibinfo{year}{2009}).

\bibitem[{\citenamefont{Butler and Kino}(1963)}]{Butler_1963}
\bibinfo{author}{\bibfnamefont{H.~S.} \bibnamefont{Butler}} \bibnamefont{and}
  \bibinfo{author}{\bibfnamefont{G.~S.} \bibnamefont{Kino}},
  \bibinfo{journal}{Phys. Fluids} \textbf{\bibinfo{volume}{6}},
  \bibinfo{pages}{1346} (\bibinfo{year}{1963}).

\bibitem[{\citenamefont{Cavaleru et~al.}(1972)\citenamefont{Cavaleru, Armour,
  and Carter}}]{Cavaleru_1972}
\bibinfo{author}{\bibfnamefont{A.~O.~R.} \bibnamefont{Cavaleru}},
  \bibinfo{author}{\bibfnamefont{D.~G.} \bibnamefont{Armour}},
  \bibnamefont{and} \bibinfo{author}{\bibfnamefont{G.}~\bibnamefont{Carter}},
  \bibinfo{journal}{Vacuum} \textbf{\bibinfo{volume}{22}}, \bibinfo{pages}{321}
  (\bibinfo{year}{1972}).

\bibitem[{\citenamefont{Demtr\"{o}der}(2008)\citenamefont{Demtr\"{o}der}}]{Demtroeder_2008}
\bibinfo{author}{\bibfnamefont{W.} \bibnamefont{Demtr\"{o}der}},
  \bibinfo{book}{\em {Laser Spectroscopy, 4th Ed.} \rm}
  (\bibinfo{publisher}{Springer},
  \bibinfo{address}{Berlin},
  \bibinfo{year}{2008}).

\bibitem[{\citenamefont{Otten}(1989)\citenamefont{Otten}}]{Otten_1989}
\bibinfo{author}{\bibfnamefont{E.W.} \bibnamefont{Otten}},
  \bibinfo{book}{\em {Investigation of Short-Lived Isotopes by Laser Spectroscopy} \rm}
  (\bibinfo{publisher}{Harwood},
  \bibinfo{address}{New York},
  \bibinfo{year}{1989}).
  
\end{thebibliography}

\end{document}